\documentclass[12pt,amsmath,onecolumn,tightenlines,nofootinbib,aps,pra]{revtex4-1}
\usepackage{graphicx}
\usepackage{hyperref}
\usepackage{bm}
\usepackage{etoolbox}
\usepackage{titlesec}
\patchcmd{\section}
{\centering}
{\raggedright}
{}
{}
\patchcmd{\subsection}
{\centering}
{\raggedright}
{}
{}
\titleformat{\section}
{\normalfont\fontsize{15}{15}\bfseries}{\thesection}{1em}{}
\renewcommand\thesection{\arabic{section}}

\newcommand{\rjp}{R(J/\psi)}
\newcommand{\jp}{J/\psi}
\newcommand{\bc}{B_c^+}

\newcommand\pubnumber{CIPANP2018-Lamm}
\newcommand\pubdate{\today}

\def\umd{Department of Physics, University of Maryland, College
Park, MD 20742, USA}
\def\support{\footnote{Work supported by the U.S.\ Department of Energy under
Contract No.\ DE-FG02-93ER-40762 and the National
Science Foundation under Grant No.\ PHY-1403891 }}

\def\Title#1{\begin{center} {\Large #1 } \end{center}}
\def\Author#1{\begin{center}{ \sc #1} \end{center}}
\def\Address#1{\begin{center}{ \it #1} \end{center}}

\newcommand\pubblock{\rightline{\begin{tabular}{l} \pubnumber\\
         \pubdate  \end{tabular}}}
\newenvironment{Abstract}{\begin{quotation}  }{\end{quotation}}
\newenvironment{Presented}{\begin{quotation} \begin{center} 
             PRESENTED AT\end{center}\bigskip 
      \begin{center}\begin{large}}{\end{large}\end{center} \end{quotation}}

\def\beq{\begin{equation}}
\def\eeq#1{\label{#1}\end{equation}}
\def\eeqn{\end{equation}}

\def\beqa{\begin{eqnarray}}
\def\eeqa#1{\label{#1}\end{eqnarray}}
\def\eeqan{\end{eqnarray}}

\let\bar=\overbar

\def\Dslash{\not{\hbox{\kern-4pt $D$}}}
\def\dslash{\not{\hbox{\kern-2pt $\del$}}}

\def\msb{{\bar{\ssstyle M \kern -1pt S}}}

\begin{document}
\begin{titlepage}
\pubblock

\vfill
\Title{Model-Independent Bounds on $R(J/\psi)$\\ via Dispersive Relations}
\vfill
\Author{ Henry Lamm\support}
\Address{\umd}
\vfill
\begin{Abstract}Model-independent bounds on $R(J/\psi) \! \equiv \!
\mathcal{BR} (B_c^+ \rightarrow J/\psi \, \tau^+\nu_\tau)/$
$\mathcal{BR} (B_c^+ \rightarrow J/\psi \, \mu^+\nu_\mu)$ are obtained through a combination of dispersive relations, heavy-quark relations at zero-recoil, and the
limited existing form factor determinations from lattice QCD\@.  The resulting
95\% confidence-level bound, $0.20\leq R(J/\psi)\leq0.39$, agrees with
the recent LHCb result at $1.3 \, \sigma$, and removes the dominant model-dependent uncertainty from theory predictions.  Using the same techniques, a prediction of $R(\eta_c)=0.29(5)$ is obtained. 
\end{Abstract}
\vfill
\begin{Presented}
Conference on the Intersections of \\Particle and Nuclear Physics\\
Palm Springs, CA, USA,  May 29-- June 30, 2018
\end{Presented}
\vfill
\end{titlepage}
\def\thefootnote{\fnsymbol{footnote}}
\setcounter{footnote}{0}
\section{Introduction}
The ratios of semileptonic heavy-meson decay branching fractions to different flavors represent sensitive tests of lepton universality and new
physics because the matrix element can be factorized at leading order into hadronic and leptonic terms:
\begin{equation}\label{eq:matel}
|\mathcal{M}_{\bar{b}\rightarrow\bar{c} \, \ell^+ \nu_\ell}|^2=
\frac{L_{\mu\nu}H^{\mu\nu}}{q^2-M_W^2}+\mathcal{O}(\alpha,G_F)\,.
\end{equation}
This expression implies that the ratios of semileptonic heavy-meson
decay branching fractions can differ from unity at this
level of precision only due to kinematic factors.  A tension between theory and experiment exist in $R(D^{(*)})$ for
heavy-light meson decays $B \! \rightarrow \!  D^{(*)}\ell\bar{\nu}$
with $\ell \! = \! \tau$ to those with $\ell \! = \! \mu$
or $e$.  In light of this tension, the LHCb Collaboration has measured the
rates for the heavy-heavy semileptonic meson decays $\bc \!
\rightarrow \! \jp \, \ell^+\nu_\ell$ in the
$\ell \! = \! \tau,\mu$ channels, finding $R(\jp) =
0.71(17)(18)$~\cite{Aaij:2017tyk}.

Prior to~\cite{Cohen:2018dgz}, only model-dependent calculations of $\rjp$ existed and although most models' central values cluster in LHCb's quoted theory
range of $0.25$--$0.28$, this range is overly optimistic, even taking only the model's own assessed uncertainty.  A more reasonable estimate of the model predictions is $0
\! < \! \rjp \! < \! 0.48$, found by forming the union of the 95\% confidence
levels (CL) using only the reported theoretical uncertainties~\cite{Cohen:2018dgz}.  Without a clear understanding
of the systematic uncertainties these assumptions introduce, even this range is suspect.

In the Standard Model, the factorization of Eq.~(\ref{eq:matel}) into
a leptonic and a hadronic tensor reduces the problem of calculating
$\rjp$ to the computation of the hadronic matrix element $\langle
\jp \, |(V \! - \! A)^\mu|\bc\rangle$.  Using this factorization,
the hadronic matrix element can be written in terms of four transition
form factors via~\cite{Wirbel:1985ji,Colquhoun:2016osw}:

\begin{align}\label{eq:hadme}
 \langle \jp(p,\epsilon)|(V-A)^\mu|\bc(P)\rangle=&
\frac{2i\epsilon^{\mu\nu\rho\sigma}}{M+m}
\epsilon^{*}_{\nu}p_{\rho}P_{\sigma}V(q^2)-(M+m)
\epsilon^{*\mu}A_1(q^2)\nonumber\\
 &+\frac{\epsilon^{*}\cdot q}{M+m}(P+p)^\mu A_2(q^2)
+2m\frac{\epsilon^{*}\cdot q}{q^2}q^{\mu}A_3(q^2)\nonumber\\
 &-2m\frac{\epsilon^{*}\cdot q}{q^2}q^{\mu}A_0(q^2) \, ,
\end{align}
 where $M \! \equiv \! M_{\bc}$ and $m \!
\equiv \!  M_{\jp}$, the momenta $P^\mu$ and
$p^\mu$, polarization $\epsilon^\mu$ of the $\jp$, and $q^\mu \! \equiv \! (P-p)^\mu$.  While five form factors are shown, only four are independent.  In the physical set,
$A_0(q^2)$ is defined as the form factor that couples to
timelike virtual $W$ polarizations ($\propto \! q^\mu$), while
$A_3(q^2)$ is simply a convenient shorthand for
\begin{equation}\label{eq:a3}
 A_3(q^2)=\frac{M+m}{2m}A_1(q^2)-\frac{M-m}{2m}A_2(q^2) \, .
\end{equation}
Furthermore, the finiteness of Eq.~(\ref{eq:hadme}) as $q^2 \! \to \!
0$ requires $A_3(0) \! = \! A_0(0)$, which is useful in
constructing bounds.  For notational simplicity, $t \! \equiv \! q^2$, and two important kinematic points are defined via $t_\pm \! = \! (M\pm m)^2$.

The state-of-the-art lattice QCD calculations for $\bc \! \rightarrow
\! \jp$ are limited to preliminary results from the HPQCD
Collaboration for $V(q^2)$ at two $q^2$ values and $A_1(q^2)$ at three
$q^2$ values~\cite{Colquhoun:2016osw} and are reproduced in Fig.~\ref{fig:latff}.  At present, there are no lattice results for $A_0(q^2)$ or $A_2(q^2)$.

While this decomposition is useful for lattice QCD, it is not the best
decomposition for the dispersive analysis.  The second convention we
use is the helicity basis, which exchanges the form factors $V,A_i$
for $g$, $f$, $\mathcal{F}_1$, and $\mathcal{F}_2$ via the relations
\begin{align}
 g=&\frac{2}{M+m} V \, , \phantom{xxx}f=(M+m)A_1 \, , \nonumber\\
 \mathcal{F}_1=\frac{1}{m}\bigg[-\frac{2k^2 t}{M+m} A_2-& \frac{1}{2}(t-M^2+m^2)(M+m)A_1\bigg] \, , \phantom{xxx}\mathcal{F}_2=2A_0 \, , \label{eq:FFrelns}
\end{align}
where $k = \sqrt{\frac{(t_+-t)(t_--t)}{4t}}$
The differential cross section for the semileptonic decay is
\begin{align}
\label{eq:difcof}
 \frac{d\Gamma}{dt}=&\frac{G_F^2|V_{cb}|^2}{192\pi^3M^3}
\frac{k}{t^{5/2}}\left(t-m_\ell^2\right)^2
\times \left\{ \left(2t+m_\ell^2\right)\left[2t|f|^2+
|\mathcal{F}_1|^2+2k^2t^2|g|^2\right] 
+3m_\ell^2k^2t|\mathcal{F}_2|^2
\right\} \, .
\end{align}

\section{Heavy-Quark Spin Symmetry}\label{sec:hqss}
In the decay $\bc \! \rightarrow \! \jp$, relations between the form factors at zero-recoil can be derived analogous to the Isgur-Wise function~\cite{Isgur:1989vq,Isgur:1989ed}.  In the heavy-heavy systems, unlike the heavy-light, the difference between the
heavy-quark kinetic energy operators produces energies no longer
negligible compared to those of the spectator $c$, and this effect
spoils the flavor symmetry in heavy-heavy systems.  Furthermore, the
spectator $c$ receives a momentum transfer from the decay of $\bar{b}
\! \to \! \bar{c}$ of the same order as the momentum imparted to the
$\bar{c}$, so one cannot justify a normalization of the form factors
at the zero-recoil point based purely upon symmetry.

While the heavy-flavor symmetry is lost, the separate spin symmetries
of $\bar{b}$ and $\bar{c}$ quarks remain, with an additional spin
symmetry from the heavy spectator $c$.  Together, the spin symmetries imply that the four form
factors are related to a single, universal function $h$ ($\Delta$ in
Ref.~\cite{Jenkins:1992nb}), but only at the zero-recoil point, and no
symmetry-based normalization for $h$ can be
derived~\cite{Jenkins:1992nb}.

In \cite{Jenkins:1992nb,Kiselev:1999sc}, the relative normalization between the four $\bar{Q}q\rightarrow \bar{Q}'q$ form
factors near the zero-recoil point [where the spatial
momentum transfer to $q$ is $\mathcal{O}(m_q)$].
The relations are:
\begin{eqnarray}\label{eq:hqss}
 g(w=1)& = & \frac{2\rho+(1+\rho)\sigma}{4M^2r\rho}f(w=1) \, ,
\nonumber \\
 \mathcal{F}_1(w=1) & = & M(1-r)f(w=1) \, , \nonumber\\
 \mathcal{F}_2(w=1) & = & \frac{2(1+r) \rho + (1-r)(1-\rho)\sigma}
{4Mr\rho}f(w=1) , \ \
\end{eqnarray}
where $r \! \equiv \! m/M$, $\rho \! \equiv \! m_{Q'} \! /m_Q$, and
$\sigma \! \equiv \! m_q/m_Q$.  These relations reproduce the standard
Isgur-Wise result~\cite{Isgur:1989vq,Isgur:1989ed,Boyd:1997kz} when
$\sigma \! = \! 0$.  The relation between
$\mathcal{F}_1(w=1)$ and $f(w=1)$ follows directly from the definition
of Eq.~(\ref{eq:FFrelns}), independent of heavy-quark symmetries.
Terms that break these relations should be $\mathcal{O}(m_c/m_b, \,
\Lambda_{\rm QCD}/m_c)\approx30\%$, and in the fits is allowed to be as large as 50\%.

\section{Dispersive Relations}\label{sec:da}

The derive constraints on the form factors of
$\bc\rightarrow\jp$ are obtained using analyticity and unitarity constraints on a
particular two-point Green's function and a conformal parameterization
in the manner implemented by Boyd, Grinstein, and Lebed
(BGL)~\cite{Grinstein:2015wqa}.  A slightly
different set of free parameters was used to simplify the the computation for the $B_c^+$ decays, and is laid out in detail in~\cite{Cohen:2018dgz,Berns:2018vpl}.

Mapping the complex $t$ plane to the unit disk
in a variable $z$ (with the two sides of the branch cut forming the
unit circle $C$) can be achieved using the conformal variable transformation
\begin{equation} \label{eq:zdef}
z(t;t_0) \equiv \frac{\sqrt{t_* - t} - \sqrt{t_* - t_0}}
{\sqrt{t_* - t} + \sqrt{t_* - t_0}} \, ,
\end{equation}
where $t_*$ is the branch point around which one deforms the contour,
and $t_0$ is a free parameter used to improve the convergence of
functions at small $z$.  In this mapping, $z$ is real for $t \le t_*$
and a pure phase for $t \ge t_*$.

To avoid issues with nonanalyticities within the unit circle, $t_* \! = \! (M_B^{(*)}-M_D)^2$, which is the lightest two particle state with the correct quantum numbers for the dispersive relations.  This is possible because for semileptonic
decays $m_\ell^2\leq t\leq t_-$ which is always less than $t_*$.  This choice ensures that the only nonanalytic features
within the unit circle $|z| \! = \! 1$ are simple poles corresponding
to single particles $B_c^{(*)+}$, which can be removed by {\it
Blaschke factors\/}~\cite{Caprini:1994fh,Caprini:1994np}. Using this formalism, each form factor $F_i(t)$ can be written as a non-analytic prefactor and an expansion
in $z$ corresponding to an analytic function.  In this way, once can show that the sum of the squares of the coefficients of $z-$expansion are bounded by one.
\section{Results}
The above constraints are summarized as:
\begin{itemize}
\item The coefficients $a_n$ of each form factor's $z-$expansion are constrained by
$\sum_n a_n^2\leq 1$.
\item Using Eq.~(\ref{eq:hqss}), the values $g(t_-)$ and
$\mathcal{F}_2(t_-)$ are related to the value of $f(t_-)$, which in
turn is computed from lattice QCD, to within 50\%.
\item All form factors are assumed maximal at the
zero-recoil point $t \! = \! t_-$ since the universal form
factor $h$ represents an overlap matrix element between initial and
final states.
\item The relation $\mathcal{F}_1(t_-)=M(1-r)f(t_-)$, which is related to the $q^2=0$ limit of the form factors, is exact.
\item $\mathcal{F}_1(0) \! = \! \frac 1 2 M^2(1-r^2)\mathcal{F}_2(0)$
follows from the condition
$A_3(0) \! = \! A_0(0)$.
\end{itemize}
 
Imposing these, the fit is performed in two steps,
reflecting the difference between the two lattice-computed form factors
($V, A_1$), and the two
($A_0, A_2$) without.
 
\begin{figure}[ht]
\includegraphics[width=0.47\linewidth]{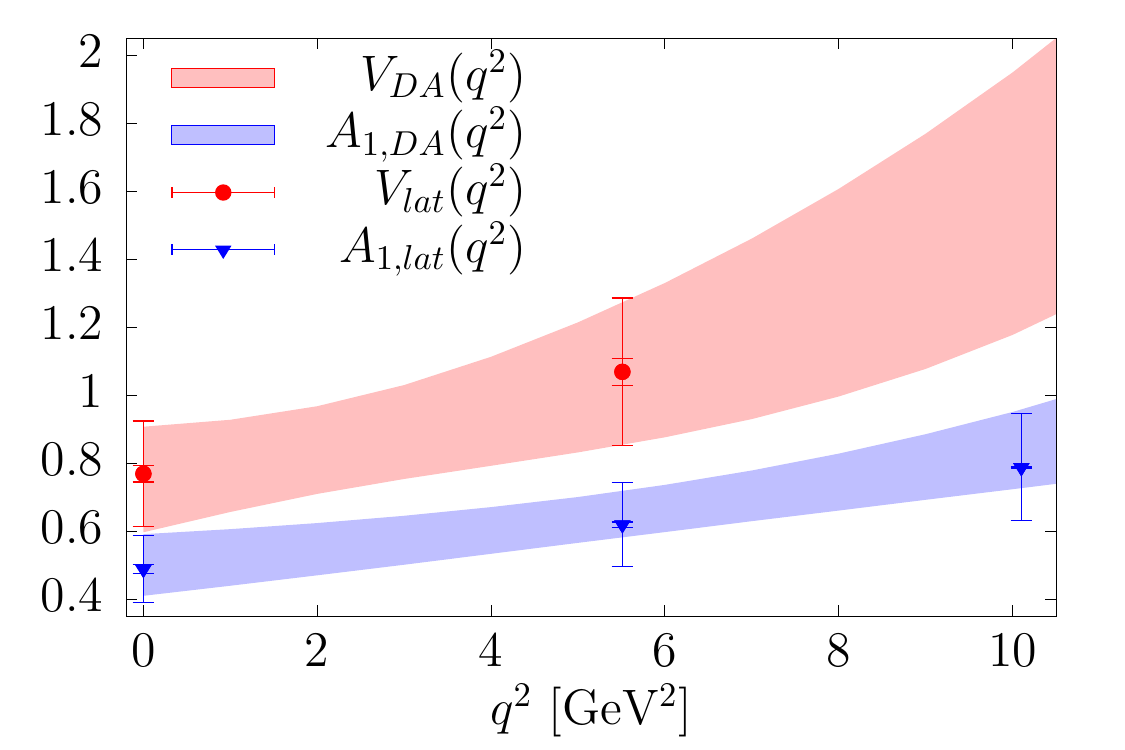}
\includegraphics[width=0.47\linewidth]{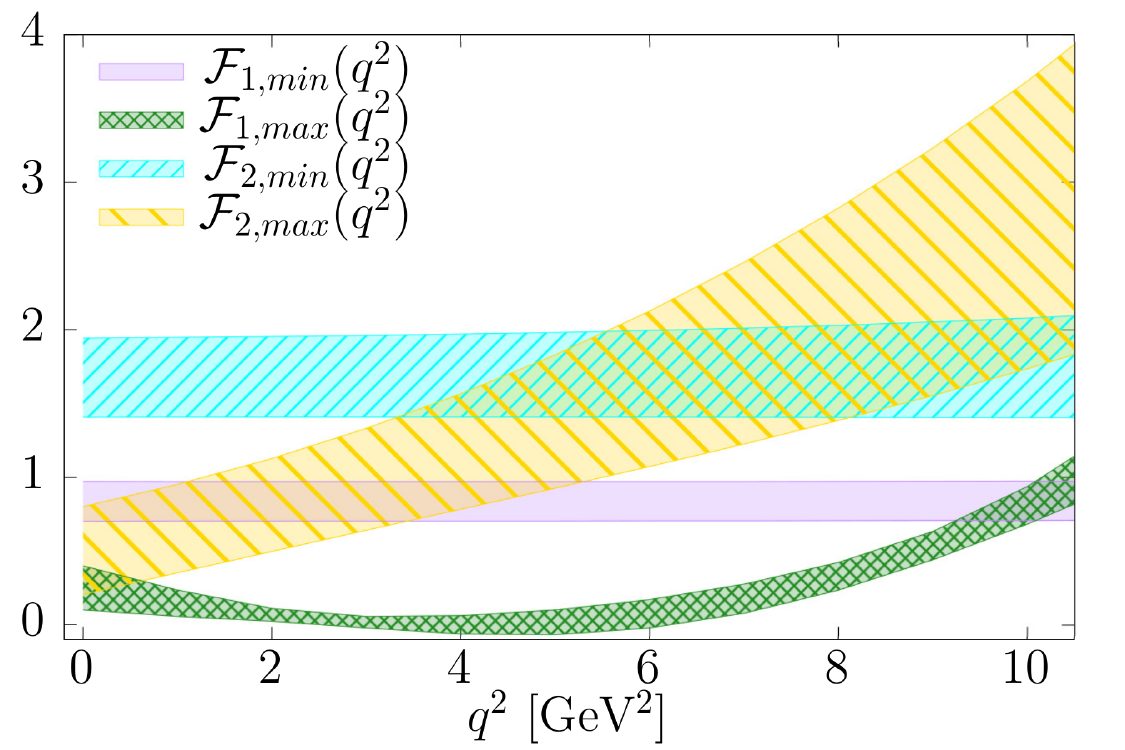}
\caption{\label{fig:latff}(left) $\bc\rightarrow \jp$ form factors $V(q^2)$
(red circles) and $A_1(q^2)$ (blue triangles) from the HPQCD
Collaboration~\cite{Colquhoun:2016osw}.  The interior bars
represent the statistical uncertainty quoted by HPQCD\@.  The exterior
bars represent the result of including our $f_{\rm lat} \! = \! 20\%$
systematic uncertainty. (right) Dimensionless form factors $\mathcal{F}_1/[
 \frac 1 2 M^2 (1 - r^2)]$ and $\mathcal{F}_2$ that provide the
 maximum and minimum $\rjp$ values consistent with lattice and
 heavy-quark spin-symmetry constraints.  In both figures, the colored bands $DA$ (dispersive analysis)
represent our one-standard-deviation ($1\sigma$) best-fit region.}
\end{figure}

In the first step, random Gaussian-distributed points are sampled for
the form factors $g$ and $f$ whose mean gives the HPQCD results and with an uncertainty dominated by an added systematic uncertainty, $f_{\rm lat}$ (expressed as a
percentage of the form-factor point value) that estimates the
uncomputed lattice uncertainties.  The resulting bands of allowed form
factors are shown for $f_{\rm lat} \! = \! 20\%$ in
Fig.~\ref{fig:latff}, alongside the HPQCD results.

In the second step, the unknown $\mathcal{F}_1$ and $\mathcal{F}_2$ are constrained. The form factors that yield the numerical maximum and minimum $\rjp$ values, subject to
the computed $f,g$ values and the constraints listed above, are obtained.  In this
way, the only uncertainties are derived from $f_lat$ and the violations of the heavy-quark spin-symmetry relations.
The resulting bands of allowed form factors that produce the minimum and maximum values of $\rjp$ are plotted in Fig.~\ref{fig:latff}. The 95\% CL ranges
for $\rjp$ as a function of the truncation power $n=1,2$ in the
dispersive analysis coefficients and $f_{\rm lat}$ are shown in
Table~\ref{tab:rvalues}\@.
\begin{table}
\caption{\label{tab:rvalues}95\% CL upper and lower bounds on
$R_{\jp}$ as a function of the truncation power $n$ of coefficients
and the systematic lattice
uncertainty $f_{\rm lat}$.}
\begin{center}
\begin{tabular}
{c c c}
\hline\hline
$f_{\rm lat}$ & $n=1$ \ & $n=2$\\
\hline
1 & [0.21, 0.33] & [0.20, 0.35]\\
5 & [0.20, 0.33] & [0.20, 0.35]\\
20 & [0.20, 0.36] & [0.20, 0.39]\\
\hline
\end{tabular}
\end{center}
\end{table}

\begin{figure}[ht]
\includegraphics[width=0.7\linewidth]{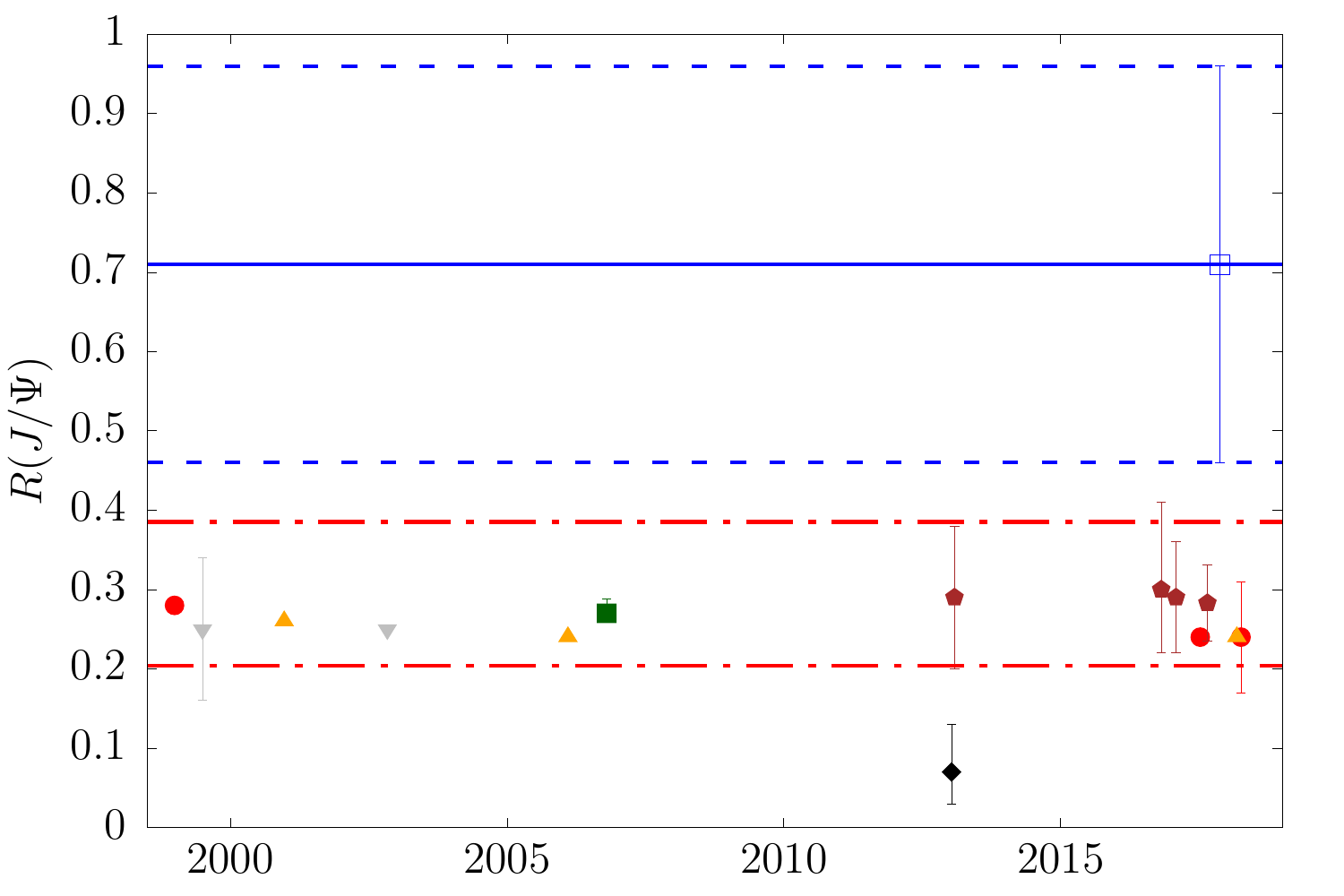}
\caption{\label{fig:rjp}$R(\jp)$ from the LHCb experiment (blue
open square, $1\sigma$ uncertainty denoted by blue dashed lines), our
bound (red dash-dotted lines), and models (points colored by model
type). Reproduced from \cite{Cohen:2018dgz}}
\end{figure}

In Fig.~\ref{fig:rjp} the previous model-dependent values of
$\rjp$ are plotted alongside the LHCb result and our 95\% CL bound of
$0.20\leq\rjp\leq0.39$, as a function of publication date.  While many of the previous model results lie within our 95\%
CL band, some are either partially or entirely excluded.

The most important piece of new lattice data to obtain is a value of $\mathcal{F}_2(t_-)$, which currently controls the upper bound's uncertainties.  This zero-recoil form factor is
directly related by $\mathcal{F}_2(t_-)=2A_0(t_-)$ to a traditional
lattice form factor.  Synthetic data suggests that a value for $\mathcal{F}_2(t_-)$ could improve the bound by the same amount as reducing the current lattice results uncertainty by a factor of 20 using far less computing resources.

The model-independent bound on $\rjp$ was found to be $0.20\leq\rjp\leq0.39$ at the 95\% CL\@.  The LHCb result is therefore consistent with the Standard
Model at $1.3 \, \sigma$.  The near-term outlook for a
higher-statistics LHCb measurement, coupled with new lattice results,
promises to reduce the experimental and theoretical uncertainties dramatically.  Using the same procedure provides a prediction of $R(\eta_c)=0.29(5)$~\cite{Berns:2018vpl} which may be obtainable in the future with LHCb.

\bibliographystyle{apsrev4-1}
\bibliography{wise}
 
\end{document}